\documentstyle[11pt,aaspp4]{article}

\slugcomment{To appear in ApJ Letters}

\lefthead{Thomsen, Baum, Hammergren, and Worthey}
\righthead{The SBF distance to Coma, and the Hubble constant}

\begin{document}

\title{The Distance to the Coma Cluster from Surface Brightness Fluctuations
       \footnotemark}

\footnotetext{Based on observations with the NASA/ESA Hubble Space Telescope
and partially supported by NASA through grant HF-1066.01-94A awarded by
the Space Telescope Science Institute, which is operated by AURA, Inc.,
under NASA Contract No. NAS5-26555.}

\author{Bjarne Thomsen}
\affil{Institute of Physics and Astronomy, University of Aarhus,
       Ny Munkegade, DK-8000 Aarhus C, Denmark; bt@obs.aau.dk}

\author{William A. Baum and Mark Hammergren}
\affil{Astronomy Department, University of Washington,
       Box 351580, Seattle, Washington 98195, USA;
       (baum,hammer)@astro.washington.edu}

\and

\author{Guy Worthey \footnotemark}
\affil{Astronomy Department, University of Michigan,
       Ann Arbor, MI 48109-1090, USA;
       worthey@astro.lsa.umich.edu}
       
\footnotetext{Hubble Fellow}

\begin{abstract}

We report on the first determination of the distance to the Coma Cluster
based on surface brightness fluctuation (SBF) measurements obtained from
{\it Hubble Space Telescope\ } WFPC2 observations of the bright E0 galaxy
NGC~4881 in the Coma Cluster and ground--based observations of the
`standard' E1 galaxy NGC~3379 in the Leo--I group. Relative distances
based on the $I$--band fluctuation magnitude, $\overline{m}_I$, are
strongly dependent on metallicity and age of the stellar population.
However, the radial changes in the stellar populations of the two
giant ellipticals, NGC~3379 and NGC~4881, are well described
by published Mg$_2$ gradients, and the ground--based measurements
of $\overline{m}_I$ at several radial points in NGC~3379 are used to
calibrate $\overline{m}_I$ in terms of the Mg$_2$ index.
The distance to NGC~3379, assumed to be identical to the average SBF
distance of the Leo--I group, is combined with the new SBF measurements
of NGC~4881 to obtain a Coma Cluster distance of $102\pm 14$~Mpc.
Combining this distance with the cosmic recession velocity of
Coma ($7186\pm428$ km s$^{-1}$), we find the Hubble constant
to be $H_0 = 71\pm 11$ km s$^{-1}$ Mpc$^{-1}$.

\end{abstract}


\section{Introduction}

To determine $H_0$, one needs to measure precise distances to galaxies
sufficiently distant for local departures from a smooth Hubble flow to
be of negligible importance. Consequently we need to extend the distance
scale to distances of order $100$ Mpc or more. Although the {\it Hubble
Space Telescope\ } ({\it HST}) has greatly extended the range of direct
Cepheid--based distances, it cannot reach the region where the Hubble
flow dominates, and despite the successful determination of many new
Cepheid distances with {\it HST}, the dispute over the global value of
the Hubble constant is still not settled.
Methods to reach the Hubble flow via the Virgo Cluster through the
use of various secondary distance indicators have tended to
obtain values larger than $70$ km s$^{-1}$ Mpc$^{-1}$
(\cite{fre94,mou95,ken95}),
whereas methods using a Cepheid calibration of the peak brightness of
type Ia supernovae consistently have given values below
$60$ km s$^{-1}$ Mpc$^{-1}$
(\cite{sah95,tam95,san96,sah96,sch96a,sch96b,bra96}).
Recently, methods based on correlations between light curve shapes
and the luminosity of type Ia supernovae (\cite{rie95,rie96,ham96}),
as well as the nebular SN Ia method based on modeling of the forbidden
line emission at late phases (\cite{rui96}), have resulted in values in
the range $60$--$70$ km s$^{-1}$ Mpc$^{-1}$.

Part of the problem is evidently the significant depth of the
Virgo Cluster (\cite{yas96}). This uncertainty is eliminated
if secondary indicators can be applied to extend distance
measurements directly to more remote clusters.
Hjorth \& Tanvir (1997) have combined the Cepheid distance to the
Leo--I group (\cite{tan95}) with a comparison of the fundamental
plane (FP) of early--type galaxies in the Leo--I group with the
FP of the Coma Cluster (\cite{jor96}).
Also, Baum {\it et al.} (1995, 1997) have directly compared the
globular cluster luminosity functions for two giant ellipticals
in the Coma Cluster with the combined luminosity function for the
Milky Way and M31. These studies also find values in the range
$60$--$70$ km s$^{-1}$ Mpc$^{-1}$. In the same spirit we shall,
for the first time, derive the distance to the Coma Cluster based on
surface brightness fluctuations.

\section{The SBF Method}

The SBF method was introduced by Tonry \& Schneider (1988)
as a precise secondary distance indicator for early--type galaxies.
Until now applications of SBF distances have been limited
to ground--based $I$--band
(\cite{ton89,ton90,ton91,jac92,ton97})
and $K$--band (\cite{lup93,pah94,jen96}) observations.
The superb resolution of the {\it HST\ } should,
however, make the WFPC2 camera an ideal instrument for obtaining
SBF distances to early--type galaxies in the Coma Cluster, the main
difficulty being the relatively modest aperture of the telescope.
A typical luminosity fluctuation must produce at least one detected
photo electron during the total exposure time in order not to be
overwhelmed by the photon noise.
The present work is based on eight $900$~s exposures of NGC~4881
obtained with the PC1 camera in the wide--band $F814W$ ($I$) filter.
The basic data reductions are fully described by Baum {\it et~al.} (1995).
Our final SBF flux $\overline{f}$ is equivalent to $\sim 0.7$ detected
photons in the $F814W$--band in a total exposure time of $7200$~s.

Surface brightness fluctuations are due to the random distribution of
discrete point sources (stars) across the field of the detector.
$\overline{f}$ is defined as the flux of a hypothetical `fluctuation star'
which, when randomly spread across the detector, produces the same flux
mean and variance per pixel as the observed stellar population.
As the statistical properties of the power spectrum are completely
determined by the mean and variance of the observed flux per pixel,
we are allowed to simulate luminosity fluctuations by a spatially random
distribution of a suitable number (in this case $\approx 1600$ stars/pixel)
of `fluctuation stars' over the field of the detector.
We have thus obtained an accurate estimate of the random error, as well
as the statistical bias, by reducing many such simulated fluctuation fields.
A high resolution Poissonian stellar fluctuation field is first convolved
with a point spread function (PSF) created by the Tiny Tim (V4.0) program
(\cite{kri93}). Subsequently, this array is rebinned to the
resolution of the PC1 camera, resulting in a noiseless stellar
fluctuation frame. Finally, Poissonian photon noise and Gaussian
read noise are added. The whole sequence is repeated $50$ times,
and all these frames are reduced in the same way as the real data.
The sample average of the derived SBF magnitudes is $0.05\pm 0.03$~mag
fainter than the true value (i.e., we must {\it subtract\ } $0.05$~mag
from the derived values), and the standard deviation on a measurement
of a single simulation ($8\times900$ s) is $0.22$~mag.
A Kolmogorov--Smirnov test (\cite{pre92}) did not detect any
deviation from normality.
Our error estimate relies completely on the Monte Carlo simulations,
which assume that the PSF is well described by the Tiny Tim model.
Fortunately, Figure~1 shows that the power spectrum of the Tiny Tim PSF
is a nearly perfect match to a high signal--to--noise power spectrum of
an annular region near the center of M32.
The power spectrum is obtained from a 400 s $F814W$ exposure with the
PC1 camera ({\it HST} archive data, ID--5464), and the fluctuation amplitude
is derived from the same angular area and by the same procedure as
we use for the NGC~4881 data. This shows that it is fully acceptable
to use the Tiny Tim model instead of an empirical PSF.

We have chosen to derive the fluctuation magnitude $\overline{m}_I$ for
an annular region between $2\farcs55$ and $5\farcs46$ from the
nucleus of NGC~4881. In that region, the white noise will be dominated by
photon statistics, not by the read noise ($R\lesssim5\farcs5$), and a good
fit to a model galaxy is obtained ($R\gtrsim2\farcs5$).
This letter is not the place for a detailed description of the SBF method,
especially as many of the reduction procedures are quite similar
to the ones described by Sodemann \& Thomsen (1995). Instead, we
shall concentrate our description on the steps that are specific
for this particular application.

The SBF amplitude, $P_0 \equiv \overline{f}$, is obtained from the radial power
spectrum of the normalized residual image and the radial power spectrum of
a high resolution PSF (a factor of 4 sub--sampling) as calculated by the
Tiny Tim (V4.0) program (\cite{kri93}). The volume of the PSF is
normalized to one, so the derived SBF amplitude (flux) corresponds
to a total magnitude. Specifically, $P_0$ is calculated using
equation (15) in Sodemann \& Thomsen (1995).
We must emphasize that the lowest wave numbers are omitted,
as they are influenced by flat--fielding and galaxy fitting errors.

In Figure~2 we plot the scaled power spectrum of the PSF on top 
of the radial power spectrum for the annular region in NGC~4881.
The constant power spectrum of the white noise, 0.030 DN,
has been subtracted from the plotted data points. Note that the
high power values with wave numbers less than 12 are not used in
our derivation of $P_0$.
We wish to thank the anonymous referee for pointing out that the
power spectrum of the NGC 4881 data appeared more noisy than the
power spectrum of the simulated data, i.e., the power spectrum of
the white noise in the NGC~4881 data was significantly higher
than expected from photon noise and assumed read noise ($5.2 e^-$).
Several identifiable sources evidently contributed to this excess 
white noise: the actual read noise ($6.2 e^-$) was higher than had
been assumed, there was a Poissonian noise associated with the 
subtracted dark current ($5.1 e^-$), and the effective number of 
frames was slightly less than 8 due to the rejection of pixels 
containing cosmic ray events.  However, these identifiable sources 
did not completely explain all of the excess white noise. Some of 
it must have been introduced by the pipeline processing.

In order to evaluate the importance of the excess white noise, we 
added the missing amount of Gaussian noise to 25 of the already 
existing simulated fluctuation frames. The standard deviation on a 
measurement of a single simulation increased from $0.22$~mag to 
$0.28$~mag. Figure~3 shows a plot of the scaled power spectrum of 
the PSF on top of the radial power spectrum for one of the simulated 
fluctuation fields with the correct amount of Gaussian noise added.
Although greater noise reduces the precision of our SBF distance
to the Coma Cluster, it does not alter the SBF distance itself.

Our calculated amplitude is
$\overline{f} = 0.0140~{\rm DN}$ in $900$~s, which according to equation
(7) of Holtzman {\it et~al.} (1995) gives a total fluctuation magnitude
of $\overline{m}_{F814W}(\infty) = 33.61$. We must {\it add\ }
$0.10$~mag to this value in order to bring it onto the instrumental
aperture ($0\farcs5$) system defined by Holtzman {\it et~al.} (1995).
In addition we must {\it add\ } $0.05$~mag because at the time of the
NGC~4881 observations the CCD was operated at $-76$~$^\circ$C instead
of $-88$~$^\circ$C, which was used after 23 April 1994 (\cite{hol95}).
We can transform the instrumental magnitude $\overline{m}_{F814W}$ to
Cousins $\overline{m}_I$ by applying Holtzman {\it et~al.}'s (1995)
synthetic transformation, which is equivalent to
$(\overline{m}_I - \overline{m}_{F814W}) = - 0.05$
(\cite[Fig. 9]{hol95}) for a typical red fluctuation color of
$(\overline{m}_V - \overline{m}_I) \approx 2.6$.
Although it is possible that the transformation from the $F814W$ band to
the Cousins $I$--band may be different for the fluctuations than it is for
stars, we derived $\overline{m}_{F814W}(\infty) = 22.90$ for a similar
region in M32, which is consistent with the transformation in light of
the $\overline{m}_I$ values given for M32 by Tonry {\it et~al.} (1990)
and Sodemann \& Thomsen (1996).
The fact that the $\overline{m}_{F814W}$ for M32
(where $(\overline{m}_V - \overline{m}_I) \approx 2.3$),
transformed according to Holtzman {\it et~al.}, agrees with the
ground--based $\overline{m}_I$ for this galaxy does not completely
ensure that there will be similar good agreement
for the redder stellar populations of giant ellipticals.
In order to settle this question we have compared
$(\overline{m}_{I}-\overline{m}_{F814W})$,
as derived from stellar population models (\cite{wor94}) having solar
metallicity and ages between 2 Gyr and 17 Gyr, with $(I-F814W)$ for the
giant stars used in the models. The fluctuation magnitude
$\overline{m}_{F814W}$ transforms, within a few hundreds of a magnitude,
in the same way as the F814W magnitude of giant stars if one chooses
the stars red enough to match the SBF colors.
Actually, a 12 Gyr model, which matches the average ${\rm Mg}_2$
index of our annulus in NGC~4881, has an offset of
$(\overline{m}_{I}-\overline{m}_{F814W}) = -0.16$ and a fluctuation color
of $(\overline{m}_{V}-\overline{m}_{I}) = 2.75$. A red giant library--star
with $(V-I) = 2.79$ has an offset of $(I-F814W) = -0.16$.
The difference between this value and the offset
$(I - F814W) = -0.05$ given by Holtzman {\it et~al.} (1995)
is almost certainly due to differences in the $I$ passbands
used to derive the synthetic colors.
Holtzman {\it et~al.} used Landolt $UBVRI$ response curves,
whereas Worthey (1994) applied the $UBVRI$ curves of Bessell (1990).
The difference between Landolt \& Bessell filter passbands
($\approx 0.1$ mag) is an indication of the remaining uncertainty
in the transformation of $F814W$ for M stars. In the following we
shall use $(\overline{m}_{I}-\overline{m}_{F814W}) = -0.10$.

The variance per pixel due to unmasked globular clusters (GCs) can be
written as $\sigma_{\rm GC}^2 = Q_{\rm GC}n_{\rm GC}f_{\rm GC}^2$
(\cite{sod95}), where $n_{\rm GC}$ is the
number of detected GCs per pixel, $f_{\rm GC}$ is a flux defining
the location of the apparent GC luminosity function (GCLF), and
$Q_{\rm GC}$ is a factor that is calculated from the apparent GCLF
and the completeness function (a detection probability).
For a Gaussian LF, $Q_{\rm GC}$ is given by equation (4) of Sodemann
\& Thomsen (1995). Following the arguments of Baum {\it et~al.}
(1995, 1997) we shall use a model based on transporting the MW+M31
GCLF to a Coma distance of $(m-M)_0 = 35.2$ (\cite{bau95}).
The combined GC data for the Milky Way and M31 are well fitted by a
hyperbolic GCLF of the form (\cite{bau97})
$\log n/n_0=-(a/b)\sqrt{b^2+(m-m_{\rm GC})^2}+k (m-m_{\rm GC})$,
where $m_{\rm GC}$ is the magnitude of the flux, $f_{\rm GC}$, at the
intersection of the asymptotes. As the detection probabilities were
derived for the $V$--band the completeness function had to be shifted
by the mean color $(V-I) = 0.97$ (\cite{bau95})
of the GC system in NGC~4881.

In a similar way, the variance per pixel due to unmasked background
galaxies can be expressed as
$\sigma_{\rm g}^2 = Q_{\rm g}n_{\rm g}f_{\rm GC}^2$,
where $n_{\rm g}$ is the number of detected background galaxies
per pixel and $Q_{\rm g}$ is a factor that is calculated from the
apparent galaxy LF and the completeness function. The slope of the
galaxy LF is taken as the average of the values given for the
$I$--band by Tyson (1988) and Lilly {\it et~al.} (1991), respectively.
For galaxies it is not so obvious that we can use our shifted
completeness function for the $V$--band to calculate $Q_{\rm g}$
for the $I$--band, because galaxies have a much larger color
spread than GCs. Fortunately, the relative contribution from
galaxies, as compared to the one from GCs, is only
$\sigma_{\rm g}^2/\sigma_{\rm GC}^2=
n_{\rm g}Q_{\rm g}/(n_{\rm GC}Q_{\rm GC}) \approx 0.05$
so we can safely neglect the error introduced by assuming that
background galaxies have the same colors as the GCs.

The contribution to the SBF flux from unmasked GCs and
background galaxies is calculated as $\Delta\overline{f} =
(Q_{\rm GC}n_{\rm GC}+Q_{\rm g}n_{\rm g})f_{\rm GC}^2/\overline{g}$,
where $\overline{g}$ is the mean galaxy flux per pixel.
There are 13 detected GCs inside the annulus [3\arcsec, 5\arcsec],
and we expect only about $0.5$ background galaxies brighter than
detection threshold within the same annulus (\cite{bau95}).
Using these values we find that we must decrease the measured SBF flux
by $\Delta\overline{f} = (2.26\pm 0.60) 10^{-3}$~DN. We must accordingly
{\it add\ } $0.19\pm 0.06$~mag in order to correct for the undetected
point sources. If we instead use the Gaussian GCLF ($V_{\rm GC} = 27.8$,
$\sigma = 1.4$) suggested by Whitmore (1996) we obtain a correction
of $0.17\pm 0.05$~mag, which is not significantly different from the
value derived for the hyperbolic GCLF. 

The K--correction for the integrated spectrum of an elliptical galaxy at
the Coma distance is quite modest for the $I$--band, but fluctuation
colors are very red, so it is much larger for $\overline{m}_I$.
The K--correction for $\overline{m}_{F814W}$ is
$K(\overline{m}_{F814W}) = (7.1\pm0.3) z = 0.16\pm0.01$ mag, where
the redshift ($z = 0.0224$) of NGC~4881 relative to the Sun
(\cite{dev91}) has been used. The quoted error was estimated from
a comparison of the K--corrections derived from stellar population
models (\cite{wor94}) having solar metallicity and ages between 5~Gyr
and 12~Gyr. The Galactic absorption in the direction of NGC~4881 is
$A_I = 0.02$ (\cite{bur84}). The final corrected fluctuation
magnitude for NGC~4881 is then
$\overline{m}_I(4881) = 33.62\pm0.29$.

\section{The Distance to NGC~4881}

We shall calibrate the SBF distance to NGC~4881 in terms of the distance
$d_{3379}=10.7\pm0.3$~Mpc (random errors) to NGC~3379 $\equiv$ M~105
in the Leo--I group. We have adopted the {\it average} SBF distance for
the Leo--I group, as recently derived by Tonry {\it et~al.} (1997)
based on the Cepheid distance scale. Sakai {\it et~al.} (1997) obtained
a direct determination of the distance
($d_{3379}=11.5$ Mpc $\pm0.7$ (random errors) $\pm1.2$ (systematic errors))
to NGC~3379 from the tip of the red--giant branch (TRGB), but the
systematic errors are relatively large.
The spiral galaxy M~96 interacts with a giant 200~kpc diameter H{\sc i}
ring (\cite{sch89}) orbiting the two central galaxies in the Leo--I group,
the E1 galaxy NGC~3379 and the S0 galaxy NGC~3384. Thus the distances to
NGC~3379 and M~96 are tied together by a physical connection.
Tanvir {\it et~al.} (1995) derived a Cepheid distance
($d_{\rm M96}=11.6\pm0.9$~Mpc) to M~96 which is very similar to the TRGB
distance obtained for NGC~3379. A significantly shorter Cepheid distance
($d_{\rm M95}=10.0\pm0.3$~Mpc) has been derived for M~95 (\cite{gra97}),
but it is not yet clear whether M~95 is at the same distance as NGC~3379.

The prototype of a normal elliptical galaxy, NGC~3379, has both a
well determined radial ${\rm Mg}_2$ gradient
${\rm Mg}_2=(0.238\pm0.002)-(0.065\pm0.003)\log({R/R_e})$ where
$R_e=56\arcsec$ (\cite{dav93}), and reliable measurements of the
$I$--band fluctuation magnitude $\overline{m}_I$ at four radial
distances between $14\farcs7$ and 48\farcs5 (\cite[Table 5]{sod95}).
This is important because $\overline{m}_I$ is very sensitive to radial
differences in metallicity and/or age of the stellar population
(\cite{ton91,ajh94,sod95,ton97}) as indicated by differences in,
e.g., the ${\rm Mg}_2$ index. As shown by a plot of $\overline{m}_I$
against ($B-I$) (\cite[Fig. 7]{sod95}) the innermost point,
being the most difficult to measure, deviates significantly
from the expected linear relation. Using a linear regression on
$\overline{m}_I$ as a function of ${\rm Mg}_2$, excluding the innermost
point, we find that
$\overline{m}_I=(28.77\pm0.03)+(7.9\pm0.7)({\rm Mg}_2-0.26)$
for the stellar population in NGC~3379. The assigned error on the
constant term mainly reflects the photometric zero point error.
This formula should only be used to interpolate $\overline{m}_I$ in
the metallicity range $0.24\lesssim {\rm Mg}_2\lesssim 0.28$. 

Fortunately, Thomsen \& Baum (1987) have measured the magnesium
gradient for NGC~4881 to be $\Delta{\rm Mg}_2/\Delta R^{1/4} =
-0.095\pm 0.027$ mag arcsec$^{-1/4}$. Combining this gradient
with Dressler's (1984) measurement ${\rm Mg}_2=0.292\pm 0.007$ of
the ${\rm Mg}_2$ index within a $4\arcsec\times 4\arcsec$ region
centered on the nucleus of NGC~4881 we obtain the following
relation: ${\rm Mg}_2 = (0.292\pm 0.007)-(0.095\pm 0.027) 
(\langle R^{1/4}\rangle-\langle R_0^{1/4}\rangle)$, where the mean
of $R^{1/4}$ for a disk with the same area as Dressler's slit is
$\langle R_0^{1/4}\rangle=1.089$, and the mean of $R^{1/4}$ for
the annulus between $2\farcs 55$ and $5\farcs 46$ is
$\langle R^{1/4}\rangle=1.424$. Our best estimate of the ${\rm Mg}_2$
index for the SBF annulus in NGC~4881 is thus ${\rm Mg}_2=0.260\pm 0.011$.
Using our relation between $\overline{m}_I$ and ${\rm Mg}_2$, a
K--correction of $K(\overline{m}_I) = 0.02$, and a Galactic extinction of
$A_I=0.02$ (\cite{bur84}), we find that a stellar population in NGC~3379
with the same ${\rm Mg}_2$ index as the one in our SBF annulus in
NGC~4881 has a fluctuation magnitude of
$\overline{m}_I(3379)=28.73\pm 0.09$.

Based on our SBF measurements the best estimate of the relative distance
modulus between NGC~3379 and NGC~4881 is $\Delta (m-M)_0 = 4.89\pm 0.30$
(random errors only), corresponding to a (luminosity) distance ratio of
$d_{4881}/d_{3379}= 9.5\pm 1.3$. Using our adopted distance to NGC~3379,
$d_{3379}=10.7\pm0.3$~Mpc, we finally obtain a distance modulus for
the Coma Cluster of $(m-M)_0 = 35.04\pm 0.31$ (random errors only),
corresponding to a luminosity distance of $d_{4881}= 102\pm 14$~Mpc. 

\section{The Hubble Constant}

At small redshifts the distance modulus is given by
$(m-M)_0 = 25 + 5 \log (c z/H_0) + \Delta m(z)$,
where $\Delta m(z) = 1.086(1-q_0)z$ (\cite{hec42}),
$z = 0.024$ is the redshift of the Coma Cluster with respect
to the cosmic background radiation (\cite{han92}), and $H_0$
is the Hubble constant in km s$^{-1}$ Mpc$^{-1}$.
If we assume that $q_0 = 0.25$ then $\Delta m(z) = 0.02$.
We adopt $c z = 7186\pm428$ km s$^{-1}$ (\cite{han92}) as likely
values for the recession and peculiar velocities of the Coma Cluster.
Finally, using these values and our distance modulus for the
Coma Cluster, $(m-M)_0 = 35.04\pm 0.31$, we estimate the Hubble
constant to be $H_0 = 71\pm 11$ km s$^{-1}$ Mpc$^{-1}$.

\acknowledgments

We are grateful to Jens Hjorth for communication of unpublished
results, and for many useful comments and discussions. We wish
to thank Marianne Sodemann for carefully reading the manuscript.

\clearpage

\begin{figure}[tinytim]
\epsscale{1.0}
\plotone{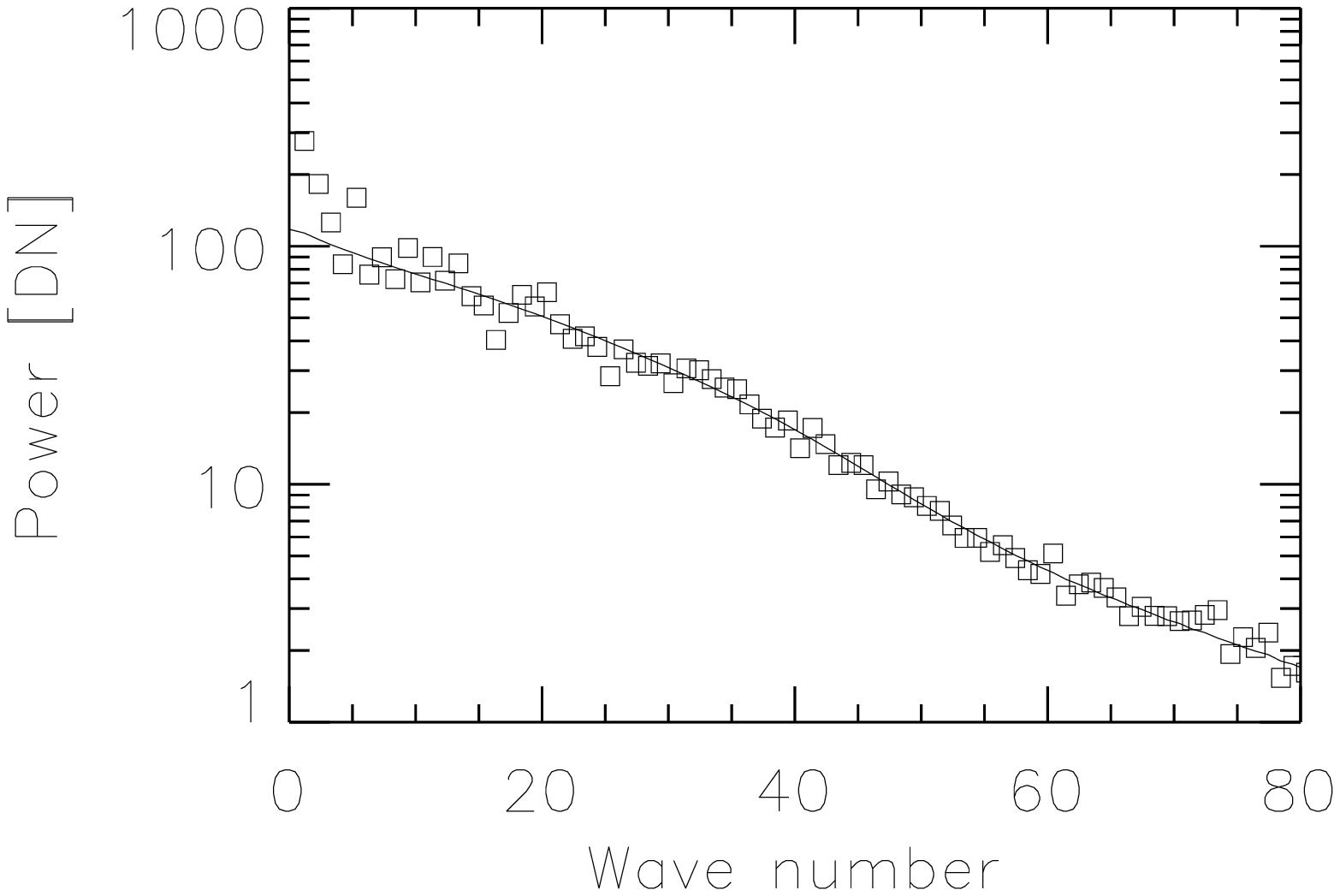}
\caption[tinytim.eps]{\label{fig1}
The scaled power spectrum of the Tiny Tim
PSF (line) is compared with the radial power spectrum of the surface
brightness fluctuations in an annular region near the center of M32
(squares). The SBF amplitude $P_0$ is derived from wave numbers
between 12 and 36 ($1.0 - 3.1$ ${\rm arcsec}^{-1}$), which is the
range used for the NGC~4881 data.
}
\end{figure}

\begin{figure}[n4881pow]
\epsscale{1.0}
\plotone{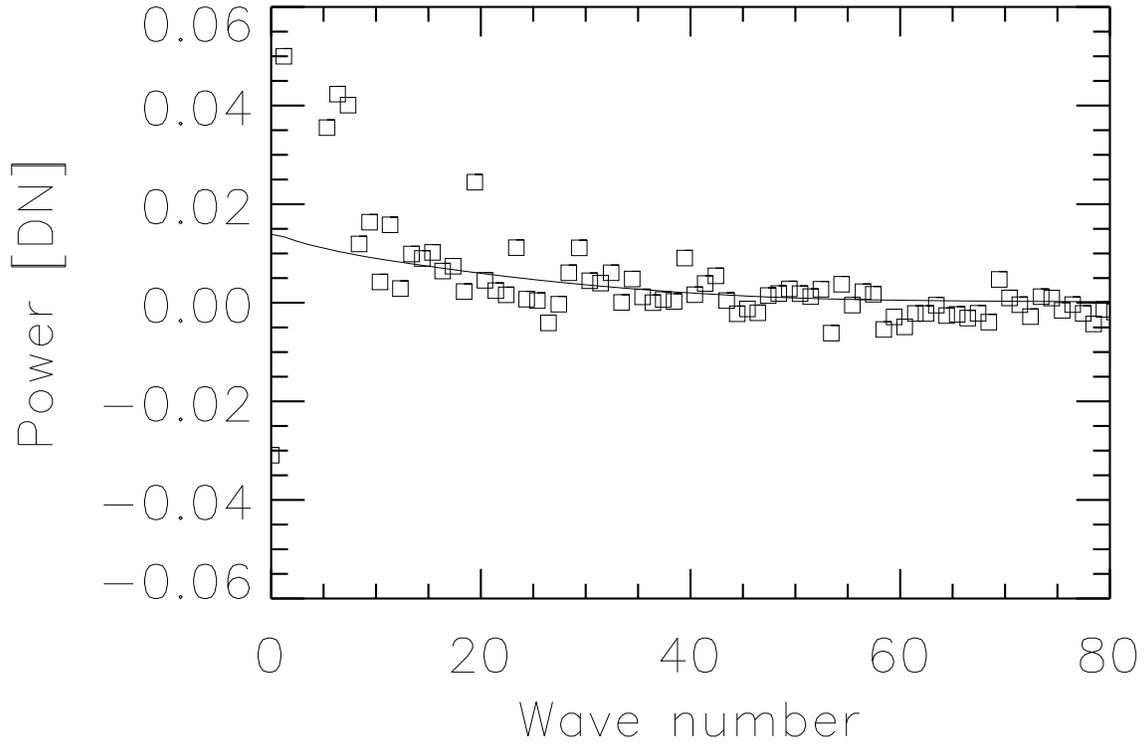}
\caption[n4881pow.eps]{\label{fig2}
The scaled power spectrum of the PSF (line)
is compared with the radial power spectrum of the surface brightness
fluctuations in the annular region [$2\farcs55$, $5\farcs46$]
centered on the nucleus of NGC~4881 (squares). The constant power
spectrum of the white noise $P_1$ (0.030 DN) has been subtracted from
the plotted data points. The SBF amplitude $P_0$ is derived from wave
numbers between 12 and 36.
}
\end{figure}

\begin{figure}[msim_pow]
\epsscale{1.0}
\plotone{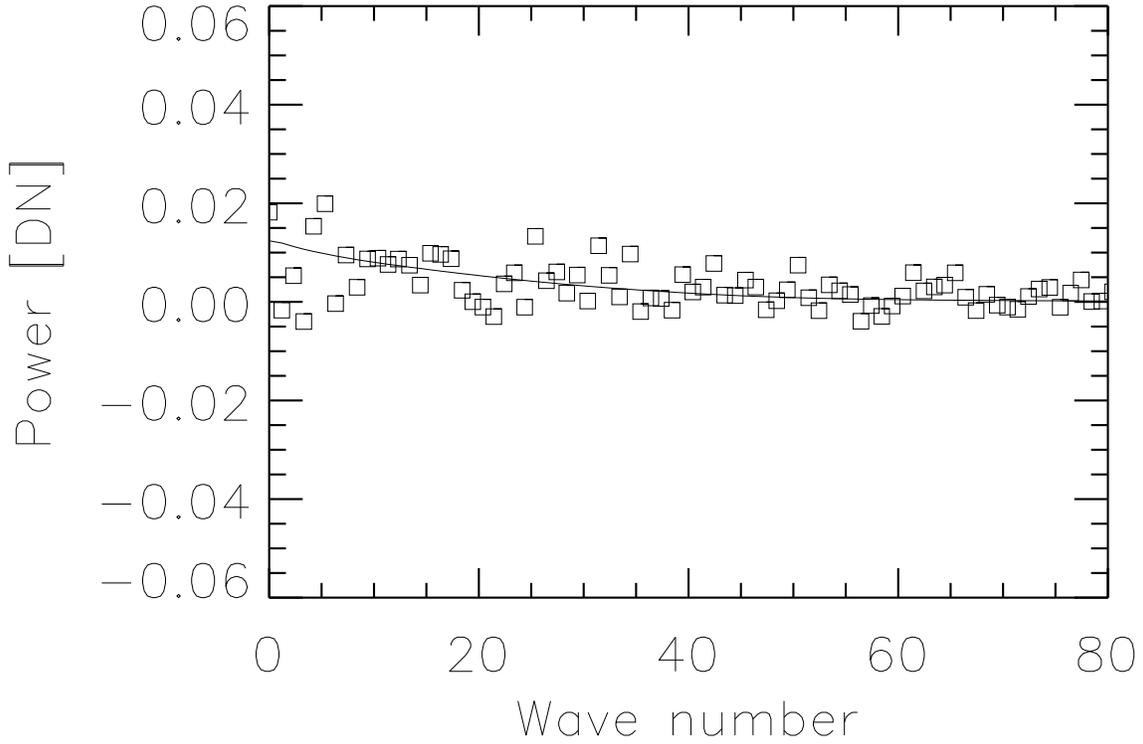}
\caption[msim_pow.eps]{\label{fig3}
The scaled power spectrum of the PSF (line)
is compared with the radial power spectrum of a simulated fluctuation
field covering the same area as used in NGC~4881 (squares).
The assumed SBF amplitude $\overline{f}$ has been reduced by the contribution
from undetected point sources. The constant power spectrum of the white
noise has been subtracted from the plotted data points. The SBF
amplitude is derived from wave numbers between 12 and 36.
}
\end{figure}


\clearpage
 
\begin{thebibliography}{}
\bibitem[Ajhar \& Tonry 1994]{ajh94} Ajhar, E. A., \& Tonry, J. L. 1994,
           \apj, 429, 557
\bibitem[Baum {\it et~al.}\ 1995]{bau95} Baum, W. A., {\it et al.} 1995, \aj,
           110, 2537
\bibitem[Baum {\it et~al.}\ 1997]{bau97} Baum, W. A., {\it et al.} 1997, \aj,
           113, in press
\bibitem[Bessell 1990]{bes90} Bessell, M. S. 1990, \pasp, 102, 1181
\bibitem[Branch {\it et~al.}\ 1996]{bra96} Branch, D.,
           Romanishin, W., \& Baron, E. 1996, \apj, 465, 73
\bibitem[Burstein \& Heiles\ 1984]{bur84} Burstein, D.,
           \& Heiles, C. 1984, \apjs, 54, 33
\bibitem[Davies {\it et~al.}\ 1993]{dav93} Davies, R. L., Sadler, E. M.,
           \& Peletier, R. F. 1993, \mnras, 262, 650
\bibitem[de Vaucouleurs {\it et~al.}\ 1991]{dev91} de Vaucouleurs, G.,
           {\it et al.} 1991,
           Third Reference Catalogue of Bright Galaxies (Springer)
\bibitem[Dressler 1984]{dre84} Dressler, A. 1984, \apj, 281, 512
\bibitem[Freedman {\it et~al.}\ 1994]{fre94} Freedman, W. L., {\it et al.}
           1994, \nat, 371, 757
\bibitem[Graham {\it et~al.}\ 1997]{gra97} Graham, J. A., {\it et al.} 1997,
           \apj, 477, 535
\bibitem[Hamuy {\it et~al.}\ 1996]{ham96} Hamuy, M., {\it et al.} 1996, \aj,
           112, 2398
\bibitem[Han \& Mould 1992]{han92} Han, M., \& Mould, J. R. 1992,
           \apj, 396, 453
\bibitem[Heckmann\ 1942]{hec42} Heckmann, O. 1942,
           in {\it Theorien der Kosmologie} (Berlin: Springer--Verlag), 63 
\bibitem[Hjorth \& Tanvir 1997]{hjo97} Hjorth, J.,
           \& Tanvir, N. R. 1997, \apj, 482, in press
\bibitem[Holtzman {\it et~al.}\ 1995]{hol95} Holtzman, J. A., {\it et al.}
           1995, \pasp, 107, 1065
\bibitem[Jacoby {\it et~al.}\ 1992]{jac92} Jacoby, G. H., {\it et al.} 1992,
           \pasp, 104, 599
\bibitem[Jensen {\it et~al.}\ 1996]{jen96} Jensen, J. B.,
           Luppino, G. A., \& Tonry, J. L. 1996, \apj, 468, 519
\bibitem[J{\o}rgensen {\it et~al.}\ 1996]{jor96} J{\o}rgensen, I.,
           Franx, M., \& Kj{\ae}rgaard, P. 1996, \mnras, 280, 167
\bibitem[Kennicutt {\it et~al.}\ 1995]{ken95} Kennicutt Jr., R. C.,
           Freedman, W. L., \& Mould, J. R. 1995, \aj, 110, 1476
\bibitem[Krist 1993]{kri93} Krist, J., \& Hasan, H. 1993, in
           {\it Astronomical Data Analysis Software and Systems II},
           ASP Conference Series 52,
           ed. R. J. Hanisch, R. J. V. Brissenden, \& J. Barnes, 530
\bibitem[Lilly {\it et~al.}\ 1991]{lil91} Lilly, S. J., Cowie, L. L.,
           \& Gardner, J. P. 1991, \apj, 369, 79
\bibitem[Luppino \& Tonry 1993]{lup93} Luppino, G. A.,
           \& Tonry, J. L. 1993, \apj, 410, 81
\bibitem[Mould {\it et~al.}\ 1995]{mou95} Mould, J., {\it et al.} 1995,
           \apj, 449, 413
\bibitem[Pahre \& Mould 1994]{pah94} Pahre, M. A., \& Mould, J. R. 1994,
           \apj, 433, 567
\bibitem[Press {\it et~al.}\ 1992]{pre92} Press, W. H., {\it et al.} 1992,
           in Numerical recipes in C. -- 2nd ed.
           (Cambridge University Press), 623
\bibitem[Riess {\it et~al.}\ 1995]{rie95} Riess, A. G.,
           Press, W. H., \& Kirshner, R. P. 1995, \apjl, 438, L17
\bibitem[Riess {\it et~al.}\ 1996]{rie96} Riess, A. G.,
           Press, W. H., \& Kirshner, R. P. 1996, \apj, 473, 88
\bibitem[Ruiz--Lapuente 1996]{rui96} Ruiz--Lapuente, P. 1996, \apjl, 465, L83
\bibitem[Saha {\it et~al.}\ 1995]{sah95} Saha, A., {\it et al.} 1995, \apj,
           438, 8
\bibitem[Saha {\it et~al.}\ 1996]{sah96} Saha, A., {\it et al.} 1996, \apj,
           466, 55
\bibitem[Sakai {\it et~al.}\ 1997]{sak97} Sakai, S., {\it et al.} 1997,
           \apj, 478, 49
\bibitem[Sandage {\it et~al.}\ 1996]{san96} Sandage, A., {\it et al.} 1996,
           \apjl, 460, L15
\bibitem[Schaefer 1996a]{sch96a} Schaefer, B. E. 1996a, \apjl, 460, L19
\bibitem[Schaefer 1996b]{sch96b} Schaefer, B. E. 1996b, \aj, 111, 1668
\bibitem[Schneider 1989]{sch89} Schneider, S. 1989, \apj, 343, 94
\bibitem[Sodemann \& Thomsen 1995]{sod95} Sodemann, M.,
           \& Thomsen, B. 1995, \aj, 110, 179
\bibitem[Sodemann \& Thomsen 1996]{sod96} Sodemann, M.,
           \& Thomsen, B. 1996, \aj, 111, 208
\bibitem[Tammann \& Sandage 1995]{tam95} Tammann, G. A.,
           \& Sandage, A. 1995, \apj, 452, 16
\bibitem[Tanvir {\it et~al.}\ 1995]{tan95} Tanvir, N. R., {\it et al.} 1995,
           \nat, 377, 27
\bibitem[Thomsen \& Baum 1987]{tho87} Thomsen, B., \& Baum, W. A. 1987,
           \apj, 315, 460
\bibitem[Tonry\ 1991]{ton91} Tonry, J. L. 1991, \apjl, 373, L1
\bibitem[Tonry {\it et~al.}\ 1989]{ton89} Tonry, J. L., Ajhar, E. A.,
           \& Luppino, G. A. 1989, \apjl, 346, L57
\bibitem[Tonry {\it et~al.}\ 1990]{ton90} Tonry, J. L., Ajhar, E. A.,
           \& Luppino, G. A. 1990, \aj, 100, 1416
\bibitem[Tonry \& Schneider 1988]{ton88} Tonry, J. L.,
           \& Schneider, D. P. 1988, \aj, 96, 807
\bibitem[Tonry {\it et~al.}\ 1997]{ton97} Tonry, J. L., Blakeslee, J. P.,
           Ajhar, E. A., \& Dressler, A. 1997, \apj, 475, 399
\bibitem[Tyson 1988]{tys88} Tyson, J. A. 1988, AJ, 96, 1
\bibitem[Whitmore 1996]{whi96} Whitmore, B. C. 1996, in
           {\it The Extragalactic Distance Scale}, STScI Preprint No. 1093
\bibitem[Worthey 1994]{wor94} Worthey, G. 1994, \apjs, 95, 107 
\bibitem[Yasuda \& Okamura 1996]{yas96} Yasuda, N., \& Okamura, S. 1996,
           \apjl, 469, L73
\end{thebibliography}
\end{document}